\documentclass[prd,twocolumn,showpacs,amsmath,amssymb,superscriptaddress,floatfix,nofootinbib]{revtex4}

\usepackage{amsfonts}
\usepackage[colorlinks=true,citecolor=blue, linkcolor=blue,urlcolor=blue]{hyperref}
\usepackage{color}
\usepackage{graphicx,amsfonts,multirow}
\usepackage{amsmath}

\begin{document}

\title{The longitudinal leading-twist distribution amplitude of $J/\psi$ meson within background field theory}

\author{Hai-Bing Fu}
\author{Long Zeng}
\address{Department of Physics, Guizhou Minzu University, Guiyang 550025, People's Republic of China}

\author{Wei Cheng}
\address{Department of Physics, Chongqing University, Chongqing 401331, People's Republic of China}

\author{Tao Zhong}
\address{Physics Department, Henan Normal University, Xinxiang 453007, People's Republic of China}

\author{Xing-Gang Wu\footnote{Corresponding author}}
\email{wuxg@cqu.edu.cn}
\address{Department of Physics, Chongqing University, Chongqing 401331, People's Republic of China}

\begin{abstract}

We make a detailed study on the $J/\psi$ meson longitudinal leading-twist distribution amplitude $\phi_{2;J/\psi}^\|$ by using the QCD sum rules within the background field theory. By keeping all the non-perturbative condensates up to dimension-six, we obtain accurate QCD sum rules for the moments $\langle\xi_{n;J/\psi}^\|\rangle$. The first three ones are $\langle\xi_{2;J/\psi}^\|\rangle=0.083(12)$, $\langle\xi_{4;J/\psi}^\|\rangle=0.015(5)$ and $\langle\xi_{6;J/\psi}^\|\rangle=0.003(2)$, leading to a single peaked behavior for $\phi_{2;J/\psi}^\|$ which is sharper than the previous ones around the region of $x\sim0.5$. As an application, we adopt the QCD light-cone sum rules to calculate the $B_c$ meson semileptonic decay $B_c^+ \to J/\psi \ell^+ \nu_\ell$. We obtain $\Gamma(B_c^+ \to J/\psi \ell^+ \nu_\ell) = (89.67^{+24.76}_{-19.06}) \times 10^{-15}~{\rm GeV}$ and $\Re(J/\psi \ell^+ \nu_\ell) = 0.217^{+0.069}_{-0.057}$, which agree with the next-to-leading order pQCD prediction and the new CDF measurement within errors.

\end{abstract}

\pacs{12.38.-t, 12.38.Bx, 14.40.Aq}

\maketitle

\section{introduction}

The $B_c^+$ meson has been discovered by the Collider Detector at Fermilab (CDF) collaboration via the semileptonic decay channel $B_c^+ \to J/\psi \ell^+ \nu_\ell$~\cite{Abe:1998wi}. At the same time, they also measured the ratio of the production cross sections times branching fractions of the $B_c^+$ meson in the decay mode $B_c^+ \to J/\psi \ell^+ \nu_\ell$ to the $B^+$ meson in the decay mode $B^+ \to J/\psi K^+$, i.e.
\begin{equation}
\Re(J/\psi \ell^+ \nu_\ell) = \frac{\sigma(B_c^+) {\cal B}(B_c^+ \to J/\psi \ell^+ \nu_\ell)}{\sigma(B^+) {\cal B}(B^+ \to J/\psi K^+)}, \label{ration}
\end{equation}
whose value was $0.132_{-0.037}^{+0.041} ({\rm st}) \pm 0.031({\rm sy}) _{-0.020}^{+0.032}({\rm lf})$, where the symbols ``st'', ``sy'' and ``lf" stand for the statistical error, the systematic error and the error of the $B_c$ meson lifetime, respectively. In year 2016, the CDF collaboration updates the value of $\Re(J/\psi \ell^+ \nu_\ell)$ by using the CDF Run II data with an integrated luminosity $8.7 ~{\rm fb}^{-1}$, i.e., $\Re(J/\psi \ell^+ \nu_\ell)=0.211\pm0.012({\rm st})^{+0.021}_{-0.020}({\rm sy})$~\cite{Aaltonen:2016dra}.

In the literature, the decay width for the semileptonic decay $B_c^+ \to J/\psi \ell^+ \nu_\ell$ has been calculated under various frameworks, such as the constituent quark model~\cite{Ivanov:2005fd, Hernandez:2006gt}, the Bethe-Salpeter (BS) equation~\cite{Chang:2001pm}, the relativistic potential model (PM)~\cite{AbdElHady:1999xh}, the perturbative QCD (pQCD) theory~\cite{Wen-Fei:2013uea}, the QCD sum rules (QCD SR)~\cite{Kiselev:1999sc, Huang:2007kb}, and the QCD light-cone sum rles (QCD LCSR)~\cite{Balitsky:1989ry, Chernyak:1990ag}. In those predictions, the decay widthes are always small, leading to large discrepancy between the experiment and the theoretical predictions on $\Re(J/\psi \ell^+ \nu_\ell)$.

Many effects have been tried to solve this discrepancy. In year 2013, a next-to-leading order (NLO) pQCD calculation gives $\Gamma(B_c^+ \to J/\psi \ell^+ \nu_\ell) = (97.30^{+36.22}_{-20.33})\times 10^{-15}~{\rm GeV}$~\cite{Qiao:2012vt}, whose accuracy has lately been improved by applying the principle of maximum conformality (PMC)~\cite{Brodsky:2011ta, Brodsky:2012rj, Mojaza:2012mf, Brodsky:2013vpa} scale-setting approach such that there is no renormalization scale independence in the decay width~\cite{Shen:2014msa}, which gives $\Gamma(B_c^+ \to J/\psi \ell^+ \nu_\ell) =(106.31^{+18.59}_{-14.01}) \times 10^{-15}~{\rm GeV}$. Those pQCD predictions lead to a larger ratio $\Re(J/\psi \ell^+ \nu_\ell)$ in agreement with the CDF run II data. However, the pQCD calculation for the $B_c \to J/\psi$ transition form factor (TFF) is only reliable in large recoil region $q^2\sim0$,  which should be extended to whole $q^2$-region via a model dependent extrapolation. This introduces extra model dependence into pQCD predictions, thus predictions from other approaches shall be helpful for a cross-check.

It has been noted that the QCD SR or LCSR TFFs are reliable for both the low and intermediate $q^2$-region, and a more reliable prediction could be expected. However, previous QCD SR or QCD LCSR predictions on those TFFs are quite small~\cite{Kiselev:1999sc, Huang:2007kb, Balitsky:1989ry, Chernyak:1990ag}, leading to a smaller $\Re(J/\psi \ell^+ \nu_\ell)$ well below the measured value. It is thus helpful to know whether the QCD SR or QCD LCSR prediction can be improved by carefully reconsidering its key components such as the TFFs and the light-cone distribution amplitudes  (LCDAs) of the $J/\psi$ meson.

The QCD LCSR is based on the operator product expansion (OPE) near the light cone $x^2 \rightsquigarrow 0$, which parameterizes all the non-perturbative dynamics into the LCDAs. Those LCDAs are non-perturbative but universal, which are key components to exclusive processes. Up to twist-4 accuracy, there are fifteen LCDAs for the vector meson, whose contributions to the $B\to\;{\rm light\; vector}$ TFFs can be grouped according to the parameter $\delta_\rho \sim m_\rho/m_B\sim 0.16$ or $\delta_{K^*} \sim m_K^*/m_B\sim0.17$~\cite{Ball:2004rg, Ball:1998sk, Ball:1998ff}. Practically, one may arrange the LCDAs of the $J/\psi$ meson following similar power counting rule, even though the parameter $\delta_{J/\psi} = m_{J/\psi}/m_{B_c}\sim 0.5$ is a little larger. As a tricky point of the LCSR approach, by choosing a proper chiral correlator, one can highlight the wanted LCDAs' contributions and greatly suppress the unwanted LCDAs' contributions to the LCSR~\cite{Huang:2001xb, Wan:2002hz, Wu:2007vi}.

By using a left-handed chiral correlator, we have calculated the LCSRs for the $B\to\rho$ TFFs in Refs.\cite{Fu:2014cna, Fu:2016yzx}. By replacing the $\rho$-meson LCDAs to those of $J/\psi$ meson, we obtain the LCSRs for the $B_c \to J/\psi$ TFFs. The resultant LCSRs for the $B_c \to J/\psi$ TFFs, similar to those of $B\to\rho$ TFFs, shall highlight the contributions from the chiral-even $J/\psi$ meson LCDAs at the $\delta_{J/\psi}^1$-order and $\delta_{J/\psi}^3$-order, which are $\phi_{2;J/\psi}^\|$, $\phi_{3;J/\psi}^\bot$, $\psi_{3;J/\psi}^\bot$, $\Phi_{3;J/\psi}^\|$, $\widetilde \Phi_{3;J/\psi}^\|$, $\phi_{4;J/\psi}^\|$ and $\psi_{4;J/\psi}^\|$, respectively. All twist-3 and twist-4 LCDAs at the $\delta_{J/\psi}^2$-order give zero contributions to the TFFs. We observe that similar to the case of $\rho$ meson, among the non-zero chiral-even LCDAs, the LCDAs $\phi_{2;J/\psi}^\|$, $\phi_{3;J/\psi}^\bot$ and $\psi_{3;J/\psi}^\bot$ provide dominant contributions to the TFFs, and the contributions from $\phi_{4;J/\psi}^\|$, $\psi_{4;J/\psi}^\|$, $\Phi_{3;J/\psi}^\|$ and $\widetilde \Phi_{3;J/\psi}^\|$ are negligible. Those LCDAs $\psi_{3;J/\psi}^\bot$ and $\phi_{3;J/\psi}^\bot$ are related to the leading-twist LCDA $\phi_{2;J/\psi}^\|$ under the Wandzura-Wilczek approximation~\cite{Ball:1997rj}. Thus, by using a left-handed chiral correlator, our main task is to determine a precise $\phi_{2;J/\psi}^\|$.

Several models for the twist-2 LCDA $\phi_{2;J/\psi}^\|$ have been suggested in the literature. For examples, Bondar and Chernyak~\cite{Bondar:2004sv}, Bodwin {\it et al.}~\cite{Bodwin:2006dm} and Sun {\it et al.}~\cite{Sun:2009zk} suggested three different models to resolve the disagreement between the experimental observations and the theoretical predictions on the production cross-section of the process $e^+ e^- \to J/\psi+ \eta_c$.

Generally, the twist-2 LCDA $\phi_{2;J/\psi}^\|$ at the scale $\mu$ can be expanded in a Gegenbauer polynomial as~\cite{Chernyak:1983ej}:
\begin{eqnarray}
\phi_{2;J/\psi}^\|(x,\mu) = 6x \bar x \left[ 1 + \sum_{n=1}^\infty a_{n;J/\psi}^\|(\mu) C^{3/2}_n(\xi) \right],  \label{HPDA_CZ}
\end{eqnarray}
where $a_{n;J/\psi}^\|(\mu)$ stands for the $n_{\rm th}$-order Gegenbauer moment. $\bar{x}=1-x$ and $\xi=x-\bar{x}$. When the scale $\mu$ is large enough, the twist-2 LCDA $\phi_{2;J/\psi}^\|(x,\mu)$ tends to the well-known asymptotic form $6x\bar x$~\cite{Lepage:1980fj}.

In the paper, we shall study the properties of $\phi_{2;J/\psi}^\|$ via studying its moments by using the Shifman-Vainshtein-Zakharov (SVZ) sum rules~\cite{Shifman:1978bx} under the background field theory (BFT). The SVZ sum rules relates the hadronic parameters, such as the meson masses and strong coupling constants, the baryon magnetic moments, and etc., to a few non-perturbative gluon and quark condensates. Those condensates are universal, and once we have determined their values by comparing with the known observables, they can be applied to all observables involving them. The SVZ sum rules approach has been applied, with remarkable success, for a large variety of properties of the low-lying hadronic states. The BFT provides a self-consistent description on those vacuum condensates and provides a systematic way to achieve the goal of the SVZ sum rules~\cite{Govaerts:1983ka, Huang:1989gv}. The SVZ sum rules for the $J/\psi$ meson LCDAs are more involved than the light vector LCDAs, since we have to take the charm-quark mass effect into consideration. Fortunately, Ref.\cite{Zhong:2014jla} gives the quark propagator and vertex operator $(z\cdot\tensor{D})^n$ with full mass dependence within the framework of BFT. Thus one can derive a precise SVZ sum rules for the moments of $\phi_{2;J/\psi}^\|$, as is the purpose of the paper.

The remaining parts of the paper are organized as follows. In Sec.\ref{section:2}, we present the SVZ sum rules for the moments of $\phi_{2;J/\psi}^\|$. Properties of the resultant $\phi_{2;J/\psi}^\|$, together with its application for the semileptonic decay $B_c^+ \to J/\psi \ell^+ \nu_\ell$, shall be discussed in Sec.\ref{section:3}. The final section is reserved for a summary.

\section{Calculation Technology}\label{section:2}

\subsection{SVZ sum rules for the moments of $\phi_{2;J/\psi}^\|$}

The QCD Lagrangian within the framework of BFT can be obtained from the conventional QCD Lagrangian by replacing the gluon field ${\cal A}_\mu^A(x)$ and quark field $\psi(x)$ to the following ones:
\begin{eqnarray}
\mathcal{A}^A_\mu(x) &\to& \mathcal{A}^A_\mu(x) + \phi^A_\mu(x), \label{bfrep0} \\
\psi(x) &\to& \psi(x) + \eta(x).
\end{eqnarray}
Here $\mathcal{A}^A_\mu(x)$ with $A =(1, \ldots, 8)$ and $\psi(x)$ are gluon and quark background fields. $\phi^A_\mu(x)$ and $\eta(x)$ are gluon and quark quantum fields, i.e., the quantum fluctuation on the background fields. The QCD Lagrangian within the BFT is given by Ref.\cite{Huang:1989gv}. The background fields satisfy the equations of motion
\begin{equation}
(i \slash \!\!\!\!  D - m)\psi(x) = 0
\end{equation}
and
\begin{equation}
\widetilde{D}^{AB}_\mu G^{B\nu\mu}(x) = g_s \bar{\psi}(x) \gamma^\nu T^A \psi(x),
\end{equation}
where $D_\mu = \partial_\mu - ig_s T^A \mathcal{A}^A_\mu(x)$ and $\widetilde{D}^{AB}_\mu = \delta^{AB} - g_s f^{ABC} \mathcal{A}^C_\mu(x)$ are fundamental and adjoint representations of the gauge covariant derivative, respectively. The physical observables should be gauge independent, one may take different gauges for the quantum fluctuations and the background fields such that to make the sum rules calculation relatively simpler. Practically, we adopt the background gauge, $\widetilde{D}^{AB}_\mu \phi^{B \mu}(x) = 0$, for the gluon quantum field~\cite{Novikov:1983gd, Hubschmid:1982pa, Govaerts:1983ka}, and the Schwinger gauge or the fixed-point gauge, $x^\mu \mathcal{A}^A_\mu(x) = 0$, for the background field~\cite{Shifman:1980ui}. Using those inputs, the quark propagator $S_F(x,0)$ and the vertex operators $\Gamma (z\cdot \tensor{D})^n$ are ready to be derived, whose explicit expressions up to dimension-six operators can be found in Ref.\cite{Zhong:2014jla}.

The twist-2 LCDA $\phi_{2;J/\psi}^\|(x,\mu)$ is defined via the following equation,
\begin{eqnarray}
&& \langle 0 | \bar{Q}_1(z) z\!\!\!\slash Q_2(-z) | J/\psi \rangle \nonumber\\
&&\qquad\qquad
= i (z\cdot q) f_{J/\psi}^\| \int^1_0 dx e^{i\xi(z\cdot q)} \phi_{2;J/\psi}^\|(x,\mu),  \label{DA_definition}
\end{eqnarray}
where $\xi = 2x-1$ and $f_{J/\psi}^\|$ is the $J/\psi$ meson decay constant. It leads to
\begin{eqnarray}
&&\langle 0|\bar Q(0) z\!\!\!\slash (iz \cdot \tensor D )^n Q(0) |J/\psi\rangle =
\nonumber\\
&&\qquad \qquad(e^{(\lambda)*} \cdot z) (q \cdot z)^n m_{J/\psi} f_{J/\psi}^\| \langle \xi_{n;J/\psi}^\| \rangle,    \label{matrix:2}
\end{eqnarray}
where $q$ and $e^{(\lambda)}$ are momentum and polarization vector of $J/\psi$ meson, $(z\cdot \tensor{D})^n = (z\cdot \overrightarrow{D} - z\cdot \overleftarrow{D})^n$. The $n_{\rm th}$-order moment $\langle \xi_{n;J/\psi}^\|\rangle$ at the scale $\mu$ is defined as
\begin{equation}
\langle \xi_{n;J/\psi}^\|\rangle = \int_0^1 dx \xi^n \phi_{2;J/\psi}^\|(x,\mu). \label{DAmoments}
\end{equation}
As a special case, the $0_{\rm th}$-moment satisfies the normalization condition
\begin{equation}
\langle \xi_{0;J/\psi}^\| \rangle = \int_0^1 dx \phi_{2;J/\psi}^\|(x,\mu) = 1 .  \label{normalizationDA}
\end{equation}

To derive the SVZ sum rules for the moments $\langle \xi_{n;J/\psi}^\|\rangle$, we introduce the following correlator,
\begin{eqnarray}
\Pi^{(n,0)}_{J/\psi} (z,q) &=& i \int d^4x e^{iq\cdot x}\langle 0|T \{ J_n(x) J^\dag_0(0)\}| 0 \rangle
\nonumber\\
&=& (z\cdot q)^{n+2} I^{(n,0)}(q^2),
\label{correlator}
\end{eqnarray}
where $J_n(x) = \bar{Q}(x) {z\!\!\!\slash} (i z\cdot \tensor{D})^n Q(x)$ and $z^2 = 0$. For the present $c\bar{c}$-system, only even moments are nonzero, $n=(0,2,4,\ldots)$.

The correlator (\ref{correlator}) is an analytic $q^2$-function. In the physical region ($q^2>0$), the hadronic content of the correlator can be quantified by inserting a complete set of the intermediate hadronic states into the matrix element with the help of the unitarity relation. By further singling out the ground-state and introducing a compact notation for the rest of contributions including excited vector mesons and continuum states, we obtain
\begin{eqnarray}
\frac{1}{\pi} \textrm{Im} I^{(n,0)}_{\rm had}(q^2) & =& \delta (q^2 - m_{J/\psi}^2) f_{J/\psi}^{\| 2} \langle \xi_{n;J/\psi}^\|\rangle  \nonumber\\
&+& \frac{3}{4\pi^2 (n+1) (n+3)} \theta (q^2 - s_{J/\psi}), \label{hadim}
\end{eqnarray}
where the quark-hadron duality has been adopted and the symbol $s_{J/\psi}$ stands for the continuum threshold for the lowest continuum state. In deep Euclidean region $q^2 < 0$, one can apply the operator product expansion for the correlator (\ref{correlator}), and the coefficients before the operators are perturbatively calculable. As a combination of the correlator within the different $q^2$-region, the sum rules for $\langle \xi^\|_{n;J/\psi}\rangle$ can be derived by using the dispersion relation. As a final step, the Borel transformation is always applied such that to suppress the contributions from excited and continuum states and those from high dimensional operators.

Following standard SVZ sum rules procedures~\cite{Shifman:1978bx, Colangelo:2000dp}, the final sum rules reads
\begin{widetext}
\begin{eqnarray}
&& \langle \xi^\|_{n;J/\psi}\rangle  =    \frac{e^{m_{J/\Psi}^2/M^2}}{f_{J/\Psi}^{\| 2}} \bigg\{ \frac{3}{8\pi^2(n + 1)(n + 3)}(1+\frac{\alpha_s}{\pi }A'_n)\int_{t_{\rm min}}^{s_{J/\psi}} ds e^{-s/M^2} \bigg[ v^{n+1} \frac{2(n+1)m_c^2+s}{s} - (v \to -v)\bigg]+ \frac{\langle {\alpha_s}G^2\rangle }{6\pi M^2}
\nonumber\\
&&\qquad\quad   \times \int_0^1 dx ~ e^{- \frac{m_c^2} {x\bar x M^2}} ~ \frac{\xi^{n-2}}{x^2\bar x^2}~\bigg[ n(n - 1)x^3 \bar x^3 +  \frac{\xi^2}{2}\bigg(1 - \frac{m^2_c (x^3 + \bar x^3)}{x^3 \bar x^3 M^2}\bigg)\bigg]+\frac{\langle g_s^3fG^3\rangle }{16\pi^2M^4}~\int_0^1 dx ~ e^{- \frac{m^2_c} {x\bar x M^2}} \frac{\xi^{n-2}}{2}~\bigg\{ \bigg[ -\xi^2
\nonumber\\
&&\qquad\quad \times\bigg(\frac{69 + 2n(11 + 64x\bar x)}{72x\bar x} + \frac{45(1 - 3x\bar x)}{8x^2\bar x^2}\bigg) - \frac{n(n - 1)}{9}[16 + (n - 31)x\bar x]\bigg] + \frac{1}{3M^2} \bigg[\xi^2\bigg(\frac{m_c^2(1+2x\bar x)}{12 x^2 \bar x^2} - \frac{8nm^2_c}{3x\bar x}
\nonumber\\
&&\qquad\quad  - \frac{3m^2_c(x^4+\bar x^4)}{4x^3\bar x^3}\bigg) + \xi \frac{11nm^2_c(x^3-\bar x^3)}{6x^2 \bar x^2} - \frac{n(n-1)m^2_c}{3}\bigg] +\xi^2  \frac{m^2_c(x^5 +\bar x^5)}{30 M^4 x^4 \bar x^4}\bigg\}\bigg\}. \label{SR_xi}
\end{eqnarray}
\end{widetext}
As an estimation of the NLO coefficients $A'_n$, we adopt the ones without quark mass effect. The first four ones are~\cite{Ball:1996tb}, $A'_0 = 1$, $A'_2 = {5}/{3}$, $A'_4 = {59}/{27}$ and  $A'_6 = {353}/{135}$, respectively. By using the moments $\langle \xi_{n;J/\psi}^\| \rangle$, we can obtain the Gegenbauer moments $a_{n;J/\psi}^\|$ by using the following relations, i.e.,
\begin{eqnarray}
 \langle \xi_{2;J/\psi}^\| \rangle &=& \frac{1}{5} + \frac{12}{35} a_{2;J/\psi}^\|, \nonumber \\
 \langle \xi_{4;J/\psi}^\| \rangle &=& \frac{3}{35} + \frac{8}{35} a_{2;J/\psi}^\| + \frac{8}{77} a_{4;J/\psi}^\|, \nonumber \\
 \langle \xi_{6;J/\psi}^\| \rangle &=& \frac{1}{21} + \frac{12}{77} a_{2;J/\psi}^\| + \frac{120}{1001} a_{4;J/\psi}^\|+\frac{64}{2145} a_{6;J/\psi}^\|, \nonumber \\
 &\cdots& \label{Eq:anxi}
\end{eqnarray}
which can be obtained by substituting Eq.(\ref{HPDA_CZ}) into Eq.(\ref{DAmoments}).

\subsection{The semi-leptonic decay for $B_c^+\to J/\psi \ell^+ \nu_\ell$}

The differential decay width for the semileptonic decay $B_c^+(P)\to J/\psi(p)\ell^+ \nu_\ell$ over the momentum transfer $q^2$ can be formulated as
\begin{widetext}
\begin{eqnarray}
&&\frac{{\rm d}\Gamma_{L}(B_c^+ \to J/\psi\ell^+ \nu_\ell)}{{\rm d}q^2} = \bigg(\frac{q^2-m_{\ell}^2}{q^2}\bigg)^2 \frac{\sqrt{\lambda(q^2)} G_F^2 |V_{\rm cb}|^2} {384m_{B_c^+}^3\pi^3} \left[\frac{3m_{\ell}^2}{q^2} \lambda(q^2) A_0^2(q^2)+(m_{\ell}^2+2q^2)|h_0(q^2)|^2 \right], \label{dGL} \\ \nonumber\\
&&\frac{{\rm d}\Gamma_{T}(B_c^+ \to J/\psi\ell^+ \nu_\ell)}{{\rm d}q^2} = \bigg(\frac{q^2-m_{\ell}^2}{q^2}\bigg)^2 \frac{\sqrt{\lambda(q^2)} G_F^2 |V_{\rm cb}|^2} {384m_{B_c^+}^3\pi^3} (m_{\ell}^2+2q^2) \left[|h_+(q^2)|^2 + |h_-(q^2)|^2 \right],\label{dGT}
\end{eqnarray}
\end{widetext}
where $q=P-p$ is the momentum transfer of the process. Here we have separated the decay width into longitudinal and transverse ones as $\Gamma=\Gamma_{L}+ \Gamma_{T}$. The lepton $\ell=e,\mu,\tau$. For the case of $\ell=e$ or $\mu$, the contributions from $A_0$ equals to zero due to the chiral suppression. The Fermi constant $G_F=1.16638\times10^{-5}$. The phase-space factor, $\lambda(q^2)=(m_{B_c^+}^2 +m_{J/\psi}^2-q^2)^2-4m_{B_c^+}^2m_{J/\psi}^2$. The longitudinal and transverse helicity amplitudes for the decay widths $\Gamma_{L}$ and $\Gamma_{T}$ are
\begin{eqnarray}
h_{\pm}(q^2) &=& \frac{\sqrt{\lambda(q^2)}}{m_{B_c^+}+m_{J/\psi}}\Big[V(q^2) \mp \frac{(m_{B_c^+}+m_{J/\psi})^2}{\sqrt{\lambda(q^2)}} \nonumber \\
&& \times A_1(q^2)\Big] , \\
h_0(q^2) &=& \frac{1}{2m_{J/\psi}\sqrt{q^2}} \Big[-\frac{\lambda(q^2)}{m_{B_c^+}+m_{J/\psi}}A_2(q^2) \nonumber\\
&+& (m_{B_c^+}+m_{J/\psi})(m_{B_c^+}^2-m_{J/\psi}^2-q^2)A_1(q^2)\Big].
\end{eqnarray}
The four $B_c\to J/\psi$ TFFs $V(q^2)$, $A_0(q^2)$, $A_1(q^2)$ and $A_2(q^2)$, as mentioned in the Introduction, can be read from Refs.\cite{Fu:2014cna, Fu:2016yzx}, which are either directly or indirectly related to the twist-2 LCDA $\phi_{2;J/\psi}^\|$.

\section{Numerical results} \label{section:3}

We adopt the following parameters to do the numerical calculation. Two non-perturbative gluon condensates are taken as $\langle\alpha_s G^2\rangle = 0.038(11)~{\rm GeV}^4$ and $\langle g_s^3 f G^3\rangle = 0.013(7)~{\rm GeV}^6$~\cite{Colangelo:2000dp, Zhong:2014jla}. The masses of $B_c^+$ and $J/\psi$ mesons are taken as $m_{B_c^+} = 6.274~{\rm GeV}$ and $m_{J/\psi} = 3.097 ~ {\rm GeV}$ from Particla data group (PDG)~\cite{Patrignani:2016xqp}. The $J/\psi$ decay constant can be related to its leptonic decay width $\Gamma_{J/\psi \to e^+e^-}$ via the relation~\cite{Hwang:1997ie}:
\begin{eqnarray}
f_{J/\psi}^{\| 2} = \frac{3}{4\pi\alpha^2 c_{J/\psi}}m_{J/\psi} \Gamma_{J/\psi \to e^+e^-},
\end{eqnarray}
where $\alpha = 1/137$ and $c_{J/\psi} = 4/9$. Taking the PDG averaged value, $\Gamma_{J/\psi \to e^+e^-} = 5.547 \pm 0.14~{\rm KeV}$~\cite{Patrignani:2016xqp}, we obtain $f_{J/\psi}^\| = 416.2 \pm 5.3~{\rm MeV}$.

\subsection{The $J/\psi$ meson leading-twist DA $\phi_{2;J/\psi}^\|(x,\mu)$}

\begin{figure}[b]
\centering
\includegraphics[width=0.45\textwidth]{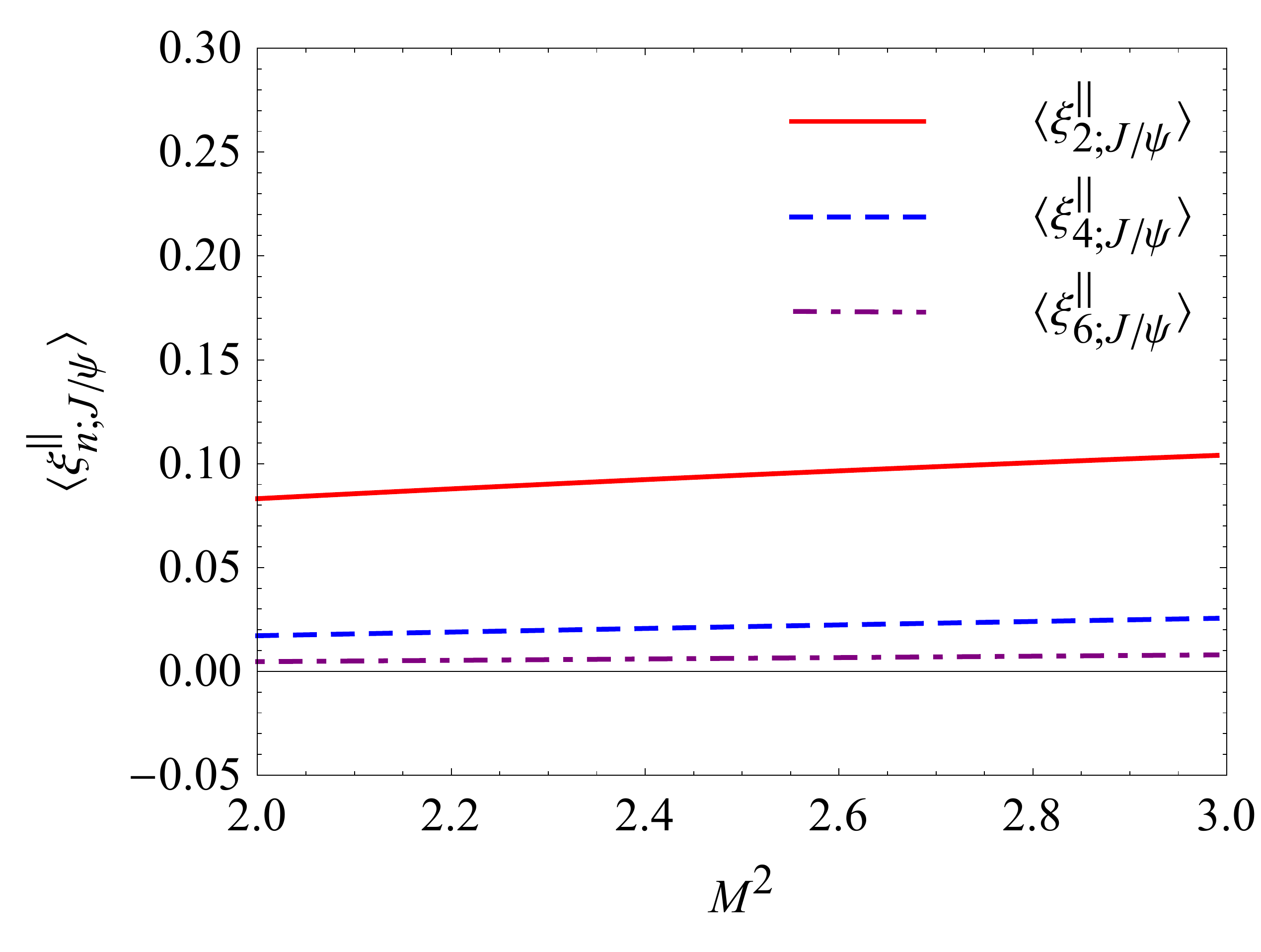}
\caption{The first three moments $\langle \xi_{n;J/\psi}^\| \rangle$ ($n=2,4,6$) versus the Borel parameter $M^2$. All input parameters are taken as their central values.} \label{Fig_xiM2}
\end{figure}

The continuum threshold $s_{J/\psi}$ for the moments $\langle \xi_{n;J/\psi}^\| \rangle$ is usually set as the value around the squared mass of the $J/\psi$ meson's first excited state. The structure of the excited $J/\psi$ meson state is not yet clear; as suggested by Braguta {\it et al}.~\cite{Braguta:2006wr}, we set the value of $s_{J/\psi}$ to infinity. To determine the Borel window, i.e. the allowable range for $M^2$, for the sum rules of the moments $\langle \xi_{n;J/\psi}^\| \rangle$, we adopt two criteria: (I) The continuum contributions are less than $40\%$ of the total dispersion relation; (II) The contributions from the dimension-six condensates should not exceed $10\%$. The determined Borel window is $M^2\in[2,3]\;{\rm GeV}^2$.  Fig.~\ref{Fig_xiM2} shows the stability of the moments $\langle \xi_{n;J/\psi}^\| \rangle$ within the allowable Borel window. It shows that the moments $\langle \xi_{4;J/\psi}^\| \rangle$ and $\langle \xi_{6;J/\psi}^\| \rangle$ are almost flat, while the value of $\langle \xi_{2;J/\psi}^\| \rangle$ changes by about $\left(^{+11\%}_{-11\%}\right)$ within the Borel window.

\begin{widetext}
\begin{center}
\begin{table}[htb]
\centering
\caption{The moments $\langle\xi_{(2,4,6);J/\psi}^\|\rangle$ of the $J/\psi$ longitudinal twist-2 DA at the scale $\mu = M$. The contributions from the LO-terms, the NLO-terms, the dimension-four and the dimension-six condensates are presented separately. The errors are squared average of all the mentioned error sources.}\label{Tab:xi_six}
\begin{tabular}{c c c c c c c c }
\hline
 ~~ ~~¡¡ & LO  & NLO & Dimension-four & Dimension-six & Total \\ \hline
$\langle \xi_{2;J/\psi}^\| \rangle  $ & 0.0890(77)& $0.0056(5)$ & $0.0001(3)$ & -0.0010(7) & 0.0937(108) \\
$\langle \xi_{4;J/\psi}^\| \rangle  $ & 0.0195(32) & $0.0016(3)$ & $0.0007(5)$ & -0.0004(2) & 0.0214(44)\\
$\langle \xi_{6;J/\psi}^\| \rangle  $ & 0.0059(14) & $0.0006(1)$ & 0.0005(3) & -0.0002(1) & 0.0063(17)\\
\hline
\end{tabular}
\label{NLO6dim}
\end{table}
\end{center}
\end{widetext}

We present the moments $\langle\xi_{(2,4,6);J/\psi}^\|\rangle$ at the scale $\mu = M$ in Table~\ref{Tab:xi_six}, where the perturbative contributions are calculated up to NLO level and the nonperturbative contributions are up to dimension-six condensates. The errors are squared averages of the uncertainties from the Borel parameter, the non-perturbative gluon condensates, and the $c$-quark mass. Table~\ref{Tab:xi_six} shows the dominant contribution is from the LO-terms, which provide $\sim95\%$ contribution to $\langle\xi_{2;J/\psi}^\|\rangle$, $\sim91\%$ to $\langle\xi_{4;J/\psi}^\|\rangle$, and  $\sim94\%$ to $\langle\xi_{6;J/\psi}^\|\rangle$, respectively. The NLO-terms provide $\sim 6.0\%$ contribution to $\langle\xi_{2;J/\psi}^\|\rangle$, $\sim 7\%$ contribution to $\langle\xi_{4;J/\psi}^\|\rangle$, and $\sim 10\%$ contribution to $\langle\xi_{6;J/\psi}^\|\rangle$. Contributions of the high dimensional condensates are small, and the condensates do not follow the usual power counting of $1/M^2$-suppression. Contribution from the dimension-six condensate has the same importance than that of the dimension-four condensate, thus both of them should be treated on an equal footing.

Using the relations among the Gegenbauer moments $a_{n;J/\psi}^\|$ and the moments $\langle \xi_{n;J/\psi}^\| \rangle$, we can get $a_{n;J/\psi}^\|$ at the same scale. The Gegenbauer moments $a_{n;J/\psi}^\|$ at any other scale can be obtained via the QCD evolution. At the NLO accuracy, we have~\cite{Floratos:1977au, Mueller:1994cn, Ball:2006nr}
\begin{eqnarray}
&&a^{\|}_{n;J/\psi}(\mu) =  a_{n;J/\psi}^\|(\mu_0) E_{n;J/\psi}^{\rm NLO} \nonumber\\
&&\qquad+ \frac{\alpha_s(\mu)}{4\pi}\sum_{k=0}^{n-2} a_{k;J/\psi}(\mu_0)\,
L^{\gamma_k^{(0)}/(2\beta_0)}d^{(1)}_{nk}.
\end{eqnarray}
Here $\mu_0$ is the initial scale, $\mu$ is the required scale, and
\begin{eqnarray}
&& E_{n;J/\psi}^{\rm NLO} =  L^{\gamma^{(0)}_n/(2\beta_0)} \nonumber\\
&&\quad \times\bigg\{1+ \frac{\gamma^{(1)}_n \beta_0-\gamma_n^{(0)}\beta_1}{8\pi\beta_0^2}
\Big[\alpha_s(\mu)-\alpha_s(\mu_0)\Big]\bigg\},
\end{eqnarray}
where $L=\alpha_s(\mu)/\alpha_s(\mu_0)$, $\beta_0=11-2n_f/3$, $\beta_1=102-38n_f/3$ with $n_f$ being the active flavor numbers, $\gamma_n^{(0)}$ and $\gamma_n^{(1)}$ are LO and NLO anomalous dimensions.

\begin{table}[b]
\centering
\caption{The moments $\langle\xi_{n;J/\psi}^\|\rangle$ at the scale $\mu_c= \bar m_c(\bar m_c)$, where the errors are squared average of all the mentioned error sources. The QCD SR prediction~\cite{Braguta:2007fh}, the Buchmuller-Tye potential model (BT model)~\cite{Buchmuller:1980su}, the Cornell potential model~\cite{Eichten:1978tg}, and the NRQCD prediction~\cite{Bodwin:2006dn} are also presented as a comparison.}
\label{Tab:xi_comparison}
\begin{tabular}{c c c c c c }
\hline
~~$\langle\xi_{n;J/\psi}^\|\rangle$~~ & ~~$n=2$~~  & ~~$n=4$~~ & ~~$n=6$~~ \\
\hline
Our prediction                    & 0.083(12) & 0.015(5) & 0.003(2) \\
QCD SR~\cite{Braguta:2007fh}       & 0.070(7)  & 0.012(2) & 0.0031(8) \\
BT model~\cite{Buchmuller:1980su} & 0.086  & 0.020 & 0.0066 \\
Cornell model~\cite{Eichten:1978tg}    & 0.084  & 0.019 & 0.0066 \\
NRQCD~\cite{Bodwin:2006dn}        & 0.075(11) & 0.010(3) & 0.0017(7) \\
\hline
\end{tabular}
\end{table}

Taking the scale as $\mu_c= \bar m_c(\bar m_c)=1.275~{\rm GeV}$ and setting other parameters to be their central values, our predictions for the moments $\langle\xi_{n;J/\psi}^\|\rangle$ are listed in Table~\ref{Tab:xi_comparison}. As a comparison, we also present the results derived within various approaches in Table~\ref{Tab:xi_comparison}, i.e. the QCD sum rules~\cite{Braguta:2007fh}, the Buchmuller-Tye potential model (BT model)~\cite{Buchmuller:1980su}, the Cornell potential model~\cite{Eichten:1978tg}, and the NRQCD~\cite{Bodwin:2006dn}. Our results agree with other predictions within errors. However, to compare with previous QCD SR prediction~\cite{Braguta:2007fh}, the central values of our $\langle\xi_{2;J/\psi}^\|\rangle$ and $\langle\xi_{4;J/\psi}^\|\rangle$ are slightly larger, leading to a sharper behavior around the region of $x\sim 0.5$ and a stronger suppression around the end point $x\sim 0,\; 1$.

\begin{figure}[htb]
\centering
\includegraphics[width=0.45\textwidth]{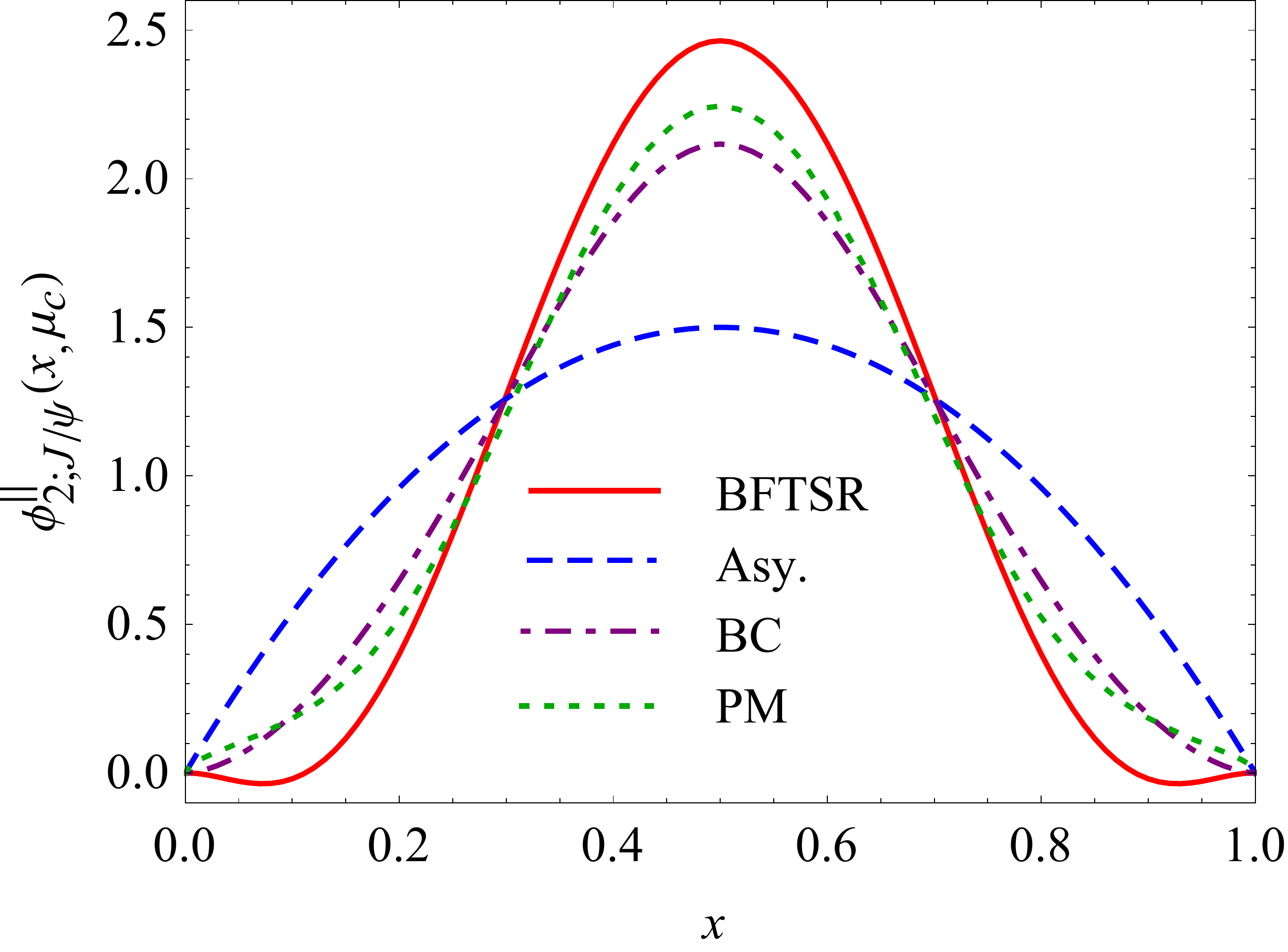}
\caption{The $J/\psi$ meson leading-twist LCDA $\phi_{2;J/\psi}^\|(x,\mu_c)$ predicted from the SVZ sum rules under the BFT (BFTSR). As a comparison, the asymptotic form, the BC model~\cite{Bondar:2004sv} and potential model~\cite{Bodwin:2006dm} are also presented. }
\label{DA:comparation}
\end{figure}

Using the relations \eqref{Eq:anxi}, the first three Gegenbauer moments at the scale $\mu_c$ are
\begin{eqnarray}
a_{2;J/\psi}^\|  &=&  -0.340(34), \\
a_{4;J/\psi}^\|  &=&  0.071(28), \\
a_{6;J/\psi}^\|  &=& 0.002(1).
\end{eqnarray}
By substituting those Gegenbauer moments into Eq.\eqref{HPDA_CZ}, we show the $J/\psi$ meson longitudinal twist-2 LCDA in Fig. \ref{DA:comparation}. As a comparison, we also present several other models in Fig. \ref{DA:comparation}, i.e. the model suggested by Bondar and Chernyak (BC)~\cite{Bondar:2004sv}:
\begin{eqnarray}
\phi_{\rm BC}(x) = c(v^2)x\bar x\bigg[\frac{x\bar x}{1 - 4x\bar x(1-v^2)}\bigg],\label{Eq:BC}
\end{eqnarray}
where $v^2=0.3$ and $c(0.3)\simeq 9.62$, the model constructed from the potential model (PM)~\cite{Bodwin:2006dm}, and the its asymptotic form $6x\bar x$. Fig. \ref{DA:comparation} shows all LCDA models prefer a single-peaked behavior, the BC and the PM LCDAs are close in shape, while our present LCDA has the sharpest peak around $x\sim0.5$ but with a strongest suppression around the end-point $x\sim0,\; 1$ \footnote{This behavior shall be helpful for suppressing the end-point singularity usually emerged in $B$-meson physics.}.

\subsection{The $B_c^+ \to J/\psi\ell^+\nu_\ell$ semileptonic decay}

One of the most important applications of the $J/\psi$ meson LCDAs is the $B_c$ meson semileptonic decay, $B_c^+ \to J/\psi \ell^+ \nu_\ell$. They are the key components of the $B_c \to J/\psi$ TFFs $A_{1}(q^2)$, $A_{2}(q^2)$ and $V(q^2)$. By using a left-handed current $j_B^\dag (x) = i\bar b(x)(1-\gamma_5)q_2(x)$ to do the LCSR calculation on the TFFs, one can suppress the contributions from other LCDAs and highlight the contributions from the longitudinal leading-twist LCDA $\phi_{2;J/\psi}^\|$, thus showing the properties of $\phi_{2;J/\psi}^\|$ via a more transparent way. Thus the LCSRs derived by using the left-handed chiral correlator~\cite{Fu:2014cna, Fu:2016yzx} inversely provide good platforms for testing the behavior of $\phi_{2;J/\psi}^\|$.

To set the Borel window for the LCSRs of the $B_c \to J/\psi$ TFFs we adopt the following criteria,
\begin{itemize}
\item We require the continuum contribution to be less than $30\%$ of the total LCSR.
\item We require all high-twist LCDAs¡¯ contributions to be less than 15\% of the total LCSR.
\item The derivatives of LCSRs for TFFs with respect to $(-1/M^2)$ give three LCSRs for the $B_c$-meson mass $m_{B_c}$. We require the predicted $B_c$-meson mass to be fulfilled in comparing with the experiment one, e.g. $|m^{\rm th}_{B_c}-m^{\rm exp}_{B_c}|/m^{\rm exp}_{B_c}$ less than 0.1\%.
\end{itemize}

\begin{table}[tb]
\centering
\caption{The TFFs at the maximum recoil point $q^2=0$. The predictions from various approaches, such as the PMC prediction~\cite{Shen:2014msa}, the QCD SR prediction with a right-handed correlator~\cite{Huang:2007kb}, the three-point sum rule (3PSR)  (with the Coloumb corrections being included)~\cite{Kiselev:1999sc}, and the quark model (QM)~\cite{Ivanov:2000aj}, are presented as a comparison. }
\label{Tab:TFF0}
\begin{tabular}{c c c c c c }
\hline
 ~~ ~~ & ~~$A_1(0)$~~  & ~~$A_2(0)$~~ & ~~$V(0)$~~ \\
 \hline
This work & $1.13^{+0.13}_{-0.11}$ & $1.20^{+0.14}_{-0.12}$ & $1.50^{+0.17}_{-0.15}$ \\
PMC~\cite{Shen:2014msa} & 1.07(52)  & 1.15(55) & 1.47(72) \\
QCD SR~\cite{Huang:2007kb} & 0.75 & 1.69 & 1.69 \\
3PSR~\cite{Kiselev:1999sc} & 0.63 & 0.69 & 1.03\\
QM~\cite{Ivanov:2000aj} & 0.68 & 0.66 & 0.96 \\
\hline
\end{tabular}
\end{table}

In agreement with previous choice of Ref.\cite{Huang:2007kb}, we take the continuum threshold for the TFFs $A_{1,2}(q^2)$ and $V(q^2)$ as $s_0 = 42.0(5)~{\rm GeV}^2$, which is smaller than the value used for LCSRs under the traditional correlator~\cite{Chabab:1993nz}. This choice in some sense ensures the contributions from the unwanted scalar resonances that are introduced by using the chiral correlator be greatly suppressed. The Borel windows are determined to be, $M^2(A_1) = 9.8(3)$, $M^2(A_2) = 11.0(3)$ and $M^2(V) = 11.0(3)$. We present the TFFs at the maximum recoil point $q^2 = 0$ in Table~\ref{Tab:TFF0}, where the errors are squared averages of all the error sources for the LCSRs. As a comparison, we also present the predictions from the NLO pQCD prediction under PMC scale-setting~\cite{Shen:2014msa}, the QCD sum rule with a right-handed correlator~\cite{Huang:2007kb}, the three-points sum rules (3PSR) (with the Coloumb corrections included)~\cite{Kiselev:1999sc}, and the quark model (QM)~\cite{Ivanov:2000aj}. It has been pointed out that the LCSRs under various choices of correlators should be consistent with each other under the same input parameters, the $B\to K^*$ TFFs are such examples~\cite{Cheng:2017bzz}. Table~\ref{Tab:TFF0} shows our LCSR predictions on the TFFs are larger than previous SR predictions, which however agrees with the PMC prediction within errors. The pQCD prediction is reliable at the maximum recoil point, thus the our present LCSR prediction could be treated as a cross-check of the NLO pQCD prediction.

\begin{figure}[tb]
\centering
\includegraphics[width=0.45\textwidth]{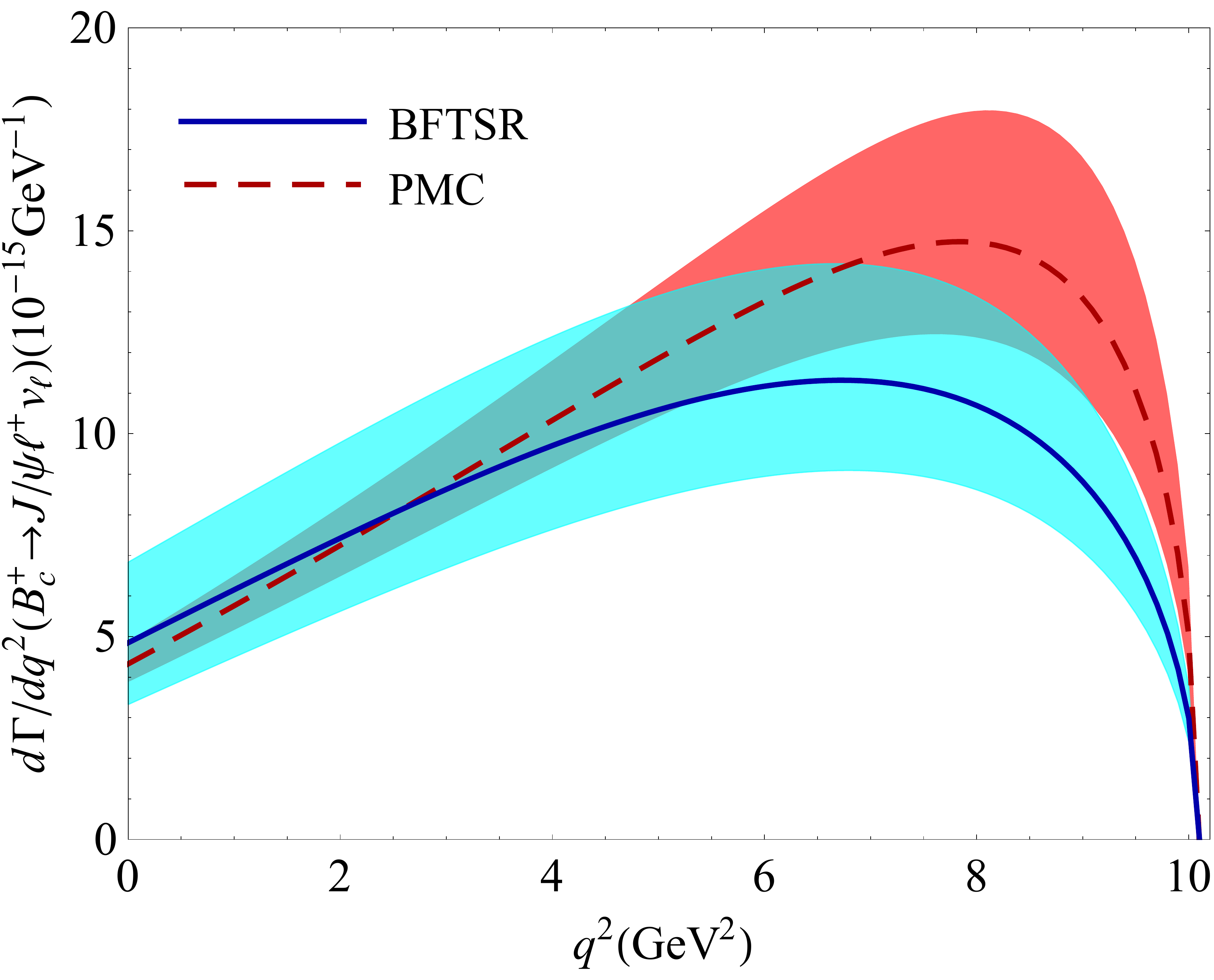}
\caption{Differential decay width for the $B_c^+ \to J/\psi\ell^+ \nu_\ell$ ($\ell = e,\mu$) versus $q^2$ by using the chiral LCSR for the TFFs and by adopting the BFTSR for the twist-2 LCDA. The PMC prediction~\cite{Shen:2014msa} is presented as a comparison.} \label{dGamma}
\end{figure}

The validity of the LCSR approach is restricted to the kinematical regime of large meson energies, and for the present case, the allowable region for $q^2$ is very close to its whole physical region, e.g. $m_\ell^2 \leq q^2 \leq (m_{B_c^+}-m_{J/\psi})^2 \approx 10~{\rm GeV}^2$. Thus we do not need to do extra extrapolations for the LCSR TFFs, while the pQCD prediction is reliable only around the maximum recoil point and certain model-dependent extrapolation has to be made, introducing extra model dependence into the pQCD prediction. We present the total differential decay width for the $B_c^+ \to J/\psi\ell^+ \nu_\ell$ ($\ell = e,\mu$) versus $q^2$ by adopting the SVZ sum rules under the BFT (BFTSR) for the twist-2 LCDA in Fig.~\ref{dGamma}, where the uncertainties are squared averages of the error sources. The PMC prediction with a monopole extrapolation~\cite{Shen:2014msa} is presented as a comparison. The BFTSR prediction agrees with the PMC prediction in low and intermediate $q^2$-region, but are smaller than the PMC one in large $q^2$-region. This difference leads to a slightly larger integrated decay width for the PMC prediction, but they are consistent with each other within reasonable errors.

\begin{table}[htb]
\centering
\caption{Total decay width (in unit: $10^{-15}~{\rm GeV}$) for the $B_c^+ \to J/\psi\ell^+ \nu_\ell$ by using the chiral LCSR for the TFFs and by adopting the BFTSR for the $J/\psi$ meson twist-2 LCDA. As a comparison, we present the results derived under various approaches, i.e. the PMC~\cite{Shen:2014msa}, the NLO pQCD calculation~\cite{Qiao:2012vt}, the QCD sum rules~\cite{Kiselev:1999sc, Huang:2007kb}, the LO pQCD calculation~\cite{Wen-Fei:2013uea}, the QCD relativistic potential model (PM)~\cite{AbdElHady:1999xh}, the Bethe-Salpeter equation~\cite{Chang:2001pm} and the constituent quark model (CQM)~\cite{Hernandez:2006gt, Ivanov:2005fd}.}
\label{tab:width}
\begin{tabular}{c c}
\hline
~~~~~~~~~~References ~~~~~~~~& ~~~~~~~~$\Gamma (B_c^+ \to J/\psi\ell^+ \nu_\ell)$~~~~~~~~\\ \hline
This work  &  $89.67^{+24.76}_{-19.06}$\\
PMC~\cite{Shen:2014msa}     & $106.31_{-14.01}^{+18.59}$ \\
NLO pQCD~\cite{Qiao:2012vt} & $97.30^{+36.22}_{-20.33}$ \\
LCSR~\cite{Huang:2007kb}       & $28\pm5$ \\
3PSR~\cite{Kiselev:1999sc}         & 34.69 \\
LO pQCD~\cite{Wen-Fei:2013uea} & $14.7^{+1.94}_{-1.73}$\\
PM~\cite{AbdElHady:1999xh}     & 30.2\\
BS equation~\cite{Chang:2001pm}        & 34.4 \\
CQM-I~\cite{Hernandez:2006gt}    & $21.9^{+1.2}$ \\
CQM-II~\cite{Ivanov:2005fd}         & 28.2 \\
\hline
\end{tabular}
\end{table}

After integrating over the allowable $q^2$-region, we get the total decay widths for $B_c^+ \to J/\psi\ell^+ \nu_\ell$ $(\ell = e,\mu)$, which are presented in Table~\ref{tab:width}. We also present the results from  the NLO pQCD prediction under PMC scale-setting~\cite{Shen:2014msa}, the NLO pQCD prediction under conventional scale-setting~\cite{Qiao:2012vt}, the QCD sum rules predictions~\cite{Huang:2007kb, Kiselev:1999sc}, the LO pQCD prediction~\cite{Wen-Fei:2013uea}, the QCD relativistic potential model prediction~\cite{AbdElHady:1999xh}, the prediction from the Bethe-Salpeter equation~\cite{Chang:2001pm}, and the constituent quark model predictions~\cite{Ivanov:2005fd, Hernandez:2006gt} in Table~\ref{tab:width}. Being consistent with Table~\ref{Tab:TFF0}, our prediction of $\Gamma{(B_c^+ \to J/\psi \ell^+ \nu_\ell)}=89.67^{+24.76}_{-19.06}\times10^{-15}~{\rm GeV}$ are about three times larger than previous SR predictions, but agree with the NLO pQCD predictions within errors. It will be found that a larger decay width is helpful for explain the large value of $\Re(J/\psi{\ell^+}\nu_\ell)$ derived from the CDF run II data~\cite{Aaltonen:2016dra}.

\begin{figure}[b]
\centering
\includegraphics[width=0.45\textwidth]{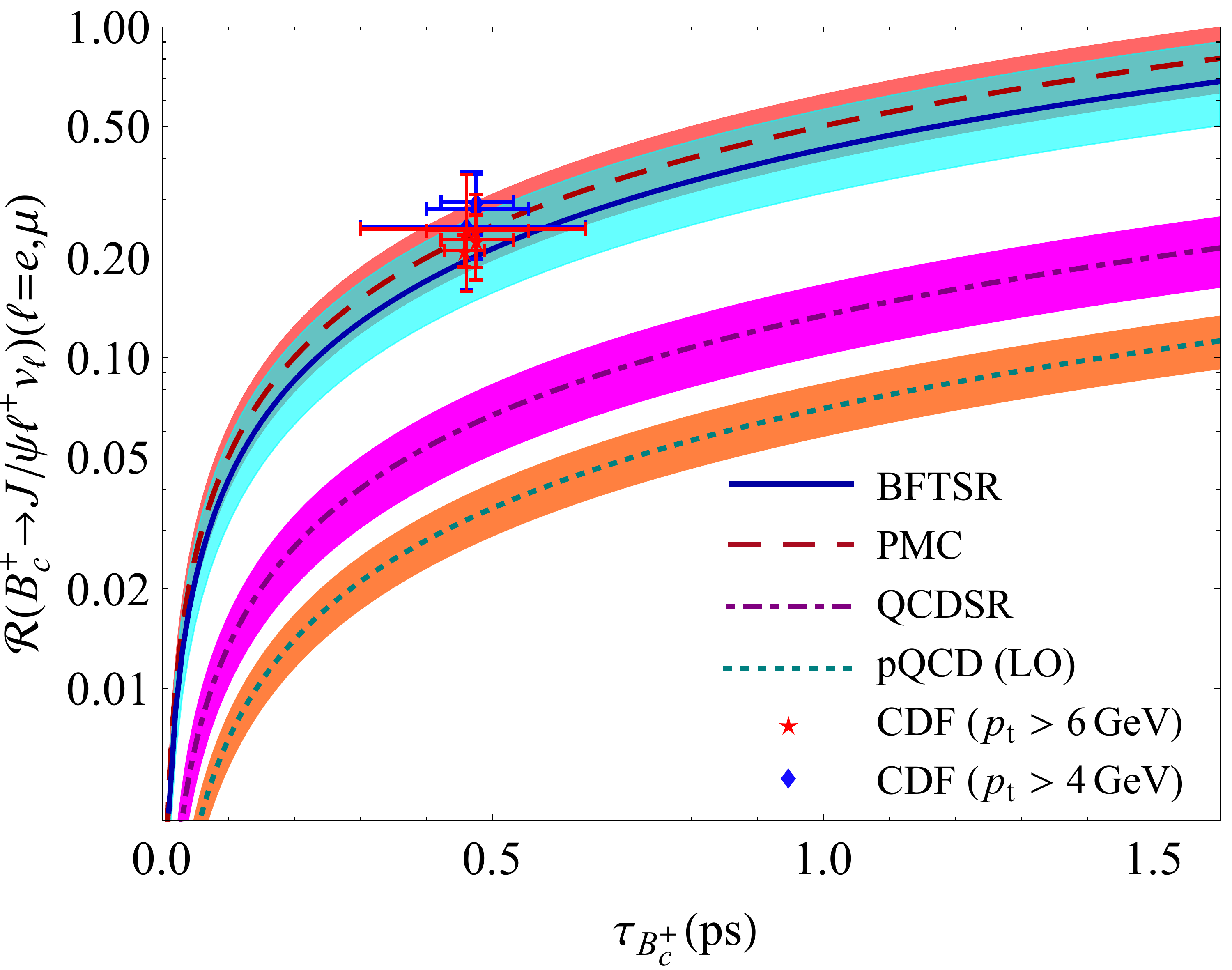}
\caption{The value of $\Re(J/\psi \ell^+\nu_\ell)$ versus the $B_c^+$ meson lifetime $\tau_{B_c^+}$ by using the chiral LCSR for the TFFs and by adopting the BFTSR for the twist-2 LCDA. The PMC~\cite{Shen:2014msa}, the QCD SR prediction~\cite{Kiselev:1999sc} and the LO pQCD prediction~\cite{Wen-Fei:2013uea} are presented as a comparison. The lines are their central values and the shaded bands are their errors. The CDF measurements~\cite{Abe:1998wi, Ru05, Ru09, Re05, Aaltonen:2016dra} are also presented. }
\label{fig:Rcompare}
\end{figure}

By further taking the hadronization fractions $f_{b\to B_c^+} = (1.3 \pm 0.2) \times 10^{-3}$, $f_{\bar b\to B^+} = 0.404\pm0.006$ and ${\cal B}(B^+ \to J/\psi K^+)=(1.026 \pm 0.031)\times 10^{-3}$~\cite{Patrignani:2016xqp}, we obtain the value of $\Re(J/\psi{\ell^+}\nu_\ell)$ defined in Eq.(\ref{ration}). The value of $\Re(J/\psi{\ell^+}\nu_\ell)$ as a function of $B_c^+$ meson lifetime $\tau_{B_c^+}$ is presented in Fig.~\ref{fig:Rcompare}. The CDF measurements~\cite{Abe:1998wi, Ru05, Ru09, Re05, Aaltonen:2016dra} as shown in Table~\ref{Tab:RR} have also been presented in Fig.~\ref{fig:Rcompare}, where all the errors are added in quadrature. All theoretical predictions on $\Re(J/\psi{\ell^+}\nu_\ell)$ are close in shape, all of which increase with the increment of $\tau_{B_c^+}$. In comparison to previous LCSR prediction such as that of Ref.\cite{Kiselev:1999sc}, our prediction of $\Re(J/\psi{\ell^+}\nu_\ell)$ shows a better agreement with the CDF measurements, being consistent with the PMC NLO pQCD prediction.

\begin{table}[tb]
\centering
\caption{Our prediction of $\sigma \cdot {\cal B}$ ratio $\Re(J/\psi{\ell^+}\nu_\ell)$. Various theoretical predictions are presented as a comparison. The CDF measurement in 2016~\cite{Aaltonen:2016dra} is also presented, where the symbols ``st'' and ``sy'' stand for the statistical error and the systematic error, respectively.} \label{Tab:RR}
\begin{tabular}{c c}
\hline
References & $\Re(J/\psi{\ell^+}\nu_\ell)$\\ \hline
This work  &  $0.217^{+0.069}_{-0.057}$\\
CDF2016~\cite{Aaltonen:2016dra} & $0.211 \pm 0.012 ({\rm st}) _{ - 0.020}^{ + 0.021} ({\rm sy})$ \\
PMC~\cite{Shen:2014msa} & $0.257^{+0.045}_{-0.034}$\\
NLO pQCD~\cite{Qiao:2012vt} & $0.235^{+0.088}_{-0.049}$\\
QCDSR-LCSR~\cite{Huang:2007kb} &    $0.068(12)$\\
QCDSR-3PSR~\cite{Kiselev:1999sc} & $0.084$\\
LO pQCD~\cite{Wen-Fei:2013uea} & $0.036^{+0.005}_{-0.004}$\\
PM~\cite{AbdElHady:1999xh}  &    $0.073$\\
BS equation~\cite{Chang:2001pm} &    $0.083$\\
CQM-I~\cite{Hernandez:2006gt} &    $0.053^{+0.003}$\\
CQM-II~\cite{Ivanov:2005fd} &   $0.068$ \\
\hline
\end{tabular}
\end{table}

If setting the $B_c^+$-meson lifetime as the PDG averaged value, $\tau_{B_c^+} = 0.507 \pm 0.009 {\rm ps}$~\cite{Patrignani:2016xqp}, we get the value of $\Re(J/\psi{\ell^+}\nu_\ell)$, which is listed in Table~\ref{Tab:RR}, in which the predictions by using the total decay width $\Gamma (B_c^+ \to J/\psi \ell^+ \nu_\ell)$ of Refs.\cite{Shen:2014msa, Qiao:2012vt, Huang:2007kb, Kiselev:1999sc, Wen-Fei:2013uea, AbdElHady:1999xh, Chang:2001pm, Hernandez:2006gt, Ivanov:2005fd} are also listed. Table~\ref{Tab:RR} shows our prediction agrees with the data issued by the CDF collaboration at year 2016~\cite{Aaltonen:2016dra}.

\section{summary}

The LCDA is an important component for QCD exclusive processes. In the paper, we make a detailed study on the $J/\psi$ longitudinal leading-twist LCDA $\phi_{2;J/\psi}^\|$ by using the QCD sum rules within the BFT. The moments of the LCDA $\phi_{2;J/\psi}^\|$ are presented in Table~\ref{Tab:xi_six}, in which the contributions from the LO-terms, the NLO-terms, the dimension-four and dimension-six operators are presented separately. It shows the LO-terms are dominant, which provide $\sim95\%$ contribution to $\langle\xi_{2;J/\psi}^\|\rangle$, $\sim91\%$ to $\langle\xi_{4;J/\psi}^\|\rangle$, and  $\sim94\%$ to $\langle\xi_{6;J/\psi}^\|\rangle$, respectively. The contribution of the dimension-four and dimension-six condensates are small, and their contribution do not follow the power counting of $1/M^2$-suppression. Thus the contribution from the dimension-six condensate has the same importance than that of the dimension-four condensate, which is also helpful for determining a more precise input parameters for the SRs. By further using the relations~\eqref{Eq:anxi}, we obtain the first three Gegenbauer moments at scale $\mu_c= \bar m_c(\bar m_c)$, $a_{2;J/\psi}^{\|}(\mu_c)  =  -0.340(34)$, $a_{4;J/\psi}^{\|}(\mu_c) =  0.071(28)$, and $a_{6;J/\psi}^{\|}(\mu_c) = 0.002(1)$.

As an application of the derived $\phi_{2;J/\psi}^\|$, we have studied the $B_c$ meson semileptonic decay $B_c^+ \to J/\psi \ell^+ \nu_\ell$. Table~\ref{Tab:TFF0} shows our LCSR predictions on the TFFs are larger than previous SR predictions, but agree with the PMC NLO pQCD prediction within errors. This leads to a larger prediction of the decay width, $\Gamma{(B_c^+ \to J/\psi \ell^+ \nu_\ell)}=89.67^{+24.76}_{-19.06}\times10^{-15}~{\rm GeV}$, which is helpful for explaining the large value of $\Re(J/\psi{\ell^+}\nu_\ell)$ obtained by the CDF run II data~\cite{Aaltonen:2016dra}, as shown explicitly by Fig.~\ref{fig:Rcompare}. If setting $\tau_{B_c^+} = 0.507 \pm 0.009 {\rm ps}$~\cite{Patrignani:2016xqp}, we obtain $\Re(J/\psi \ell^+ \nu_\ell) = 0.217^{+0.069}_{-0.057}$, which agrees well with the CDF predictions in year 2016 and the PMC NLO pQCD prediction. \\

{\bf Acknowledgments}: We are grateful to Jian-Ming Shen for helpful discussions. This work was supported in part by the Natural Science Foundation of China under Grant No.11647112, No.11625520, No.11547015 and No.11765007; by the Project of Guizhou Provincial Department of Science and Technology under Grant No.[2017]1089; by the Project for Young Talents Growth of Guizhou Provincial Department of Education under Grant No.KY[2016]156; the Key Project for Innovation Research Groups of Guizhou Provincial Department of Education under Grant No.KY[2016]028.

\end{document}